\documentclass[aps,pra,preprint,groupedaddress,showpacs]{revtex4}

\usepackage{graphicx}
\usepackage{bm}
\usepackage{amsmath,amssymb}
\usepackage{txfonts}

\begin{document}


\title{Exchange effects in elastic collisions of spin-polarized electrons 
with open-shell molecules with ${}^3 \Sigma_{g}^{-}$ symmetry
}


\author{Motomichi Tashiro}
\email[E-mail:]{tashiro@fukui.kyoto-u.ac.jp}

\affiliation{Fukui Institute for Fundamental Chemistry,
Kyoto University, Takano-Nishi-Hiraki-cho 34-4, Kyoto 606-8103, Japan.}


\date{\today}

\begin{abstract}
The spin-exchange effect in spin-polarized electron collisions with unpolarized open-shell 
molecules, O$_2$, B$_2$, S$_2$ and Si$_2$, has been studied by the R-matrix method with 
the fixed-bond approximation. All of these molecules have ${}^3 \Sigma_{g}^{-}$ symmetry 
in their ground states. 
Usual integrated cross sections with unpolarized electrons has also been studied. 
We used the complete active space self consistent 
field orbitals and put more than 10 target electronic states in the R-matrix models. 
In electron O$_2$ elastic collisions, calculated polarization fractions agree well 
with the experimental results, especially around the ${}^4 \Sigma_u^-$ resonance. 
In e-B$_2$, S$_2$ and Si$_2$ elastic collisions, larger spin-exchange effect is observed 
compared to the e-O$_2$ elastic collisions. 
In all four cases, spin-exchange effect becomes prominent near resonances.  
This association of resonance and magnitude of the spin-exchange effect was studied 
by explicitly removing the resonance configurations from the R-matrix calculations. 
In general, spin-exchange effect is larger in e-B$_2$ collisions than in e-S$_2$ and Si$_2$ collisions,  
and is smallest in e-O$_2$ collisions. 
\end{abstract}


\maketitle


\section{Introduction}

When electron collides elastically with open-shell atom or molecule, they can exchange 
their spins. Thus, spin polarization of the electron beam is 
in general reduced after scattering with unpolarized open-shell targets. 
We can obtain more precise information of the scattering process by studying this 
depolarization \cite{1987ChJPh..25..156H, 1988JPhB...21.2621B, 1993JPhB...26.4607H},  
which is difficult to observe in usual experiment with spin-averaging procedure.  


Collisions of spin-polarized electrons with atoms have been studied for long years 
(see Hegemann et al.\cite{PhysRevLett.66.2968} and references therein). 
In contrast, number of experiment on electron molecule system is limited. 
Ratliff et al.\cite{1989PhRvA..39.5584R} measured rate constants for electron exchange   
in elastic electron collisions with O$_2$ ${X}^3 \Sigma_{g}^{-}$ and NO ${X}^2 \Pi$ molecules 
in thermal energies. 
Their spin-exchange rate constants are substantially smaller than those 
in electron hydrogen-atom or alkali-metal-atom collisions. 
Hegemann et al.\cite{PhysRevLett.66.2968,1993JPhB...26.4607H} studied exchange process in 
elastic electron collisions with O$_2$ ${X}^3 \Sigma_{g}^{-}$ and NO ${X}^2 \Pi$ molecules and 
Na ${3}^2 S$ atoms. They measured ratio of spin-polarizations in electron beams before 
and after collisions, i.e., polarization fraction, which is directly related to the spin-exchange 
differential cross sections. 
As in the work of Ratliff et al.\cite{1989PhRvA..39.5584R}, Hegemann et al.\cite{PhysRevLett.66.2968} 
confirmed that the exchange cross sections of electron molecule collisions are much smaller than 
those of electron collisions with atoms. 
Although absolute value of spin-exchange cross section is small, the degree of spin-exchange in electron 
O$_2$ collisions becomes larger at 100 degrees with collision energies between 8 and 15 eV compared to 
the other angles and energies, which they attributed to the existence of 
the O$_2^-$  ${}^4 \Sigma_u^-$ resonance. 


Theoretical study of spin-exchange in electron O$_2$ collisions was performed 
by da Paixao et al.\cite{1992PhRvL..68.1698D}. They used the Schwinger 
multichannel method with the three lowest electronic states of O$_2$ 
in their model, and confirmed that spin-exchange cross section is small in electron 
O$_2$ ${X}^3 \Sigma_{g}^{-}$ elastic scatterings. 
Although exchange cross sections are small for electron collisions with randomly oriented O$_2$, 
they observed large depolarization at some scattering angles when electrons were scattered from 
spatially oriented O$_2$ molecules. The profile of depolazation as a function of scattering 
angle depends strongly on molecular orientation. 
Based on these resuls, da Paixao et al.\cite{1992PhRvL..68.1698D} explained for the first time 
that the experimental exchange cross section in electron-molecule collisions is small because 
averaging over molecular orientation washes out depolarization effects. 
Fullerton et al.\cite{1994JPhB...27..185F}, Nordbeck et al.\cite{1994JPhB...27.5375N} and 
W{\"o}ste et al.\cite{1996JPhB...29.2553W} used the R-matrix method 
to calculate polarization fractions in electron O$_2$ collisions. 
The calculations of Fullerton et al.\cite{1994JPhB...27..185F} and 
Nordbeck et al.\cite{1994JPhB...27.5375N} employed the fixed 
bond approximation with T-matrix elements obtained by the nine-state R-matrix 
calculation of Noble and Burke \cite{No92}, whereas W{\"o}ste et al.\cite{1996JPhB...29.2553W} 
used vibrational averaging of T-matrices to include the effect of nuclear motion. 
The fixed bond R-matrix calculations of Fullerton et al.\cite{1994JPhB...27..185F} and 
Nordbeck et al.\cite{1994JPhB...27.5375N} 
confirmed the results of da Paixao et al.\cite{1992PhRvL..68.1698D}. 
Agreement with experimental results at energies from 10 to 15 eV is marginal, 
as in the calculation of da Paixao et al.\cite{1992PhRvL..68.1698D}. 
The vibrational averaging procedure of W{\"o}ste et al.\cite{1996JPhB...29.2553W} 
improved agreement with the experimental results in this energy region. 
Machado et al.\cite{1999PhRvA..60.1199M} applied the Schwinger variational iterative 
method combined with the distorted-wave approximation and obtained 
similar elastic e-O$_2$ polarization fractions to those of Fullerton et al.\cite{1994JPhB...27..185F}.

Other than electron O$_2$ collisions, theoretical work of spin exchange 
in electron molecule collisions is scarce. 
da Paixao et al.\cite{PhysRevA.53.1400} calculated polarization fractions 
in electron NO ${X}^2 \Pi$ elastic collisions as they did in electron O$_2$ 
collisions \cite{1992PhRvL..68.1698D}. Calculated exchange effect was small 
in e-NO elastic collisions, in agreement with the experimental results of 
Hegemann et al.\cite{PhysRevLett.66.2968,1993JPhB...26.4607H}. 
Sartori et al.\cite{PhysRevA.55.3243} studied spin-exchange in  
the superelastic electron collisions with H$_2$ c ${}^3 \Pi_u$ state 
using the Schwinger multichannel method. Large depolarization was observed 
in their results, intermediate between depolarizations in e-Na and 
e-O$_2$ collisions. 
%
Recently, Fujimoto et al.\cite{2006PhRvA..73a2714F} performed 
the iterative Schwinger variational calculation of spin-exchange effect 
in elastic electron C$_2$O ${X}^3 \Sigma^{-}$ collisions. 
They found modest depolarization near resonances, however, 
spin-exchange effect was very small in other energy region. 
%
%



Recently, we have studied electron O$_2$ scatterings by the R-matrix method with 
improved molecular orbitals and increased number of target states\cite{2006PhRvA..73e2707T,2006PhRvA..74b2706T} 
compared to the previous theoretical studies.  
Our results are in good agreement with the previous experimental results. 
Since the previous theoretical polarization fractions of elastic e-O$_2$ collisions at 
energies between 10 and 15eV agree not so well with the experimental results, 
it would be interesting to examine how spin-exchange cross section will change in this energy region 
by our improved calculational parameters. 
At the same time, it is important to understand general behaviour of 
spin-exchange cross sections, polarization fractions in other words, 
in elastic electron molecule collisions. 
Until now, spin-exchange effect in low-energy electron molecule elastic scattering 
has been studied only for NO and C$_2$O molecules other than O$_2$. 
Thus it is desirable to study spin-exchange in other electron-molecule scattering systems as well. 
In this work, we study spin-exchange in electron O$_2$ collisions with the same 
calculational parameters as we used in our previous works\cite{2006PhRvA..73e2707T,2006PhRvA..74b2706T}. 
In addition, we calculate spin-exchange cross section in elastic electron collisions with 
B$_2$, S$_2$ and Si$_2$ molecules. These B$_2$, S$_2$ and Si$_2$ are stable homo-nuclear 
molecules with ${}^3 \Sigma_g^-$ symmetry in their ground states, as in O$_2$ molecule. 


In this paper, details of the calculations are presented in section 2, 
and we discuss the results in section 3 comparing our results with 
previous theoretical and available experiments. 
Then the summary is given in section 4. 

\section{Theoretical methods}

\subsection{polarization fraction}

In this work, we consider elastic scattering of spin-polarized electrons 
from randomly oriented unpolarized molecules. 
When the incident electrons have spin polarization $P$  
and the scattered electrons have polarization $P'$,  
with the polarization direction perpendicular to the scattering plane, 
the polarization fraction, the ratio of $P$ and $P'$, is a measure of spin-exchange and is 
related to the spin-flip differential cross section (DCS) 
$d\sigma_{\rm SF}/d\Omega$ as \cite{1993JPhB...26.4607H, 1994JPhB...27..185F},
\begin{equation}
\frac{P'}{P} = 1 - 2 \frac{d\sigma_{\rm SF}/d\Omega}{d\sigma/d\Omega}. 
\label{eq1}
\end{equation}
Here $d\sigma/d\Omega$ is the DCS obtained by unpolarized electrons. 
%

The DCSs of $d\sigma/d\Omega$ and $d\sigma_{\rm SF}/d\Omega$ are 
evaluated by the spin-specific scattering amplitude \cite{1994JPhB...27..185F, 2006PhRvA..74b2706T}, 
\begin{equation}
f^{S}_{ij} = \sum_{l_{i} m_{i} l_{j} m_{j}} \sum_{\Gamma \lambda \mu \nu} 
\frac{ \sqrt{\pi \left(2l_{i}+1\right) } }{\sqrt{k_{i}k_{j}}} 
i^{l_{i}-l_{j}+1} 
\mathcal{D}_{0 ~ \lambda}^{l_{i}~ *} \left( \alpha \beta \gamma \right)
\mathcal{D}_{\nu~\mu}^{l_{j}} \left( \alpha \beta \gamma \right)
Y_{l_{j}}^{\nu} \left( \hat{r} \right) C_{\lambda,m_{i}} C_{\mu,m_{j}}^{*}
T_{ij}^{\Gamma S M_{S}}, 
\label{eq5}
\end{equation} 
where $i$ and $j$ specify the states of the target molecule as well as the scattering electron in the 
initial and final channels, respectively.  
$k_i$ and $k_j$ are the initial and final wavenumber of the electron, 
$\mathcal{D}_{m~ m'}^{l} \left( \alpha \beta \gamma \right)$ 
is the rotation matrix with the Euler angles 
$\left(\alpha, \beta, \gamma \right)$ representing orientation of the target molecule 
in the laboratory frame. 
The electron is scattered to the direction $\hat{r}$ in the laboratory frame in 
this expression. 
The T-matrix elements $T_{ij}^{\Gamma S M_{S}}$ are prepared 
for all possible spin $S$ of the electron-molecule system as well as 
all irreducible representation $\Gamma$ of the symmetry of the 
system. We used $D_{2h}$ in the R-matrix calculations. 
Since the target molecules have triplet spin symmetry 
in this work, we only include $S=1/2$ and $3/2$ in our calculations. 
The matrix element $C_{\lambda,m}$ relates the spherical harmonics 
$Y_l^{\lambda}$ to the real spherical harmonics $S_l^m$. The explicit 
expression of $C_{\lambda,m}$ can be found in our previous paper \cite{2006PhRvA..74b2706T}. 

In this paper, we consider elastic scattering of electron from molecule 
with triplet spin symmetry. 
Then, $d\sigma/d\Omega$ and $d\sigma_{\rm SF}/d\Omega$ are expressed by 
the spin specific amplitude $f^S$ as \cite{1993JPhB...26.4607H, 1994JPhB...27..185F},

\begin{equation}
\frac{d\sigma}{d\Omega} = 
\frac{1}{3}\left( 2 \left| f^{3/2} \right|^{2} + \left| f^{1/2} \right|^{2} \right),
\label{eq2}
\end{equation}
and 
\begin{equation}
\frac{d\sigma_{\rm SF}}{d\Omega} = 
\frac{4}{27}\left| f^{3/2} - f^{1/2} \right|^{2}. 
\label{eq3}
\end{equation}
Here summations over channel indices are omitted for notational simplicity. 
Note that the expression of the spin-flip DCS contains interference 
of amplitudes with different spin multiplicities. 
Since the target molecules are randomly oriented, these DCSs are 
averaged over all possible molecular orientations in space. 


\subsection{Detail of the R-matrix calculation}

The T-matrix elements $T_{ij}^{\Gamma S M_{S}}$ 
were obtained by a modified version of the polyatomic programs in the UK 
molecular R-matrix codes \cite{Mo98}. 
General procedure of calculation is almost the same as in our 
previous works\cite{2006PhRvA..73e2707T,2006PhRvA..74b2706T}. 
Since the R-matrix method itself has been described extensively
in the literature \cite{Bu05,Go05,Mo98} and references therein, 
we do not repeat general explanation of the method here. 

In this work, elastic electron collisions with O$_2$, B$_2$, S$_2$ and Si$_2$ 
molecules were studied. For electron O$_2$ scattering, we used the same 
parameter set as we used in the previous works\cite{2006PhRvA..73e2707T,2006PhRvA..74b2706T}. 
Specifically, we employed the equilibrium bond length of 2.300 a$_0$ for O$_2$, 
the R matrix radius of 10 a$_0$. The angular quantum number of the 
scattering electron was included up to $l$=5. 
The atomic basis set for bound molecular 
orbitals, number of the target states included in the model 
as well as choice of the configurations in the inner region calculation 
were the same. 

For the electron B$_2$, S$_2$ and Si$_2$ scatterings, we included 14, 13 and 15 
target electronic states in the R-matrix calculation, respectively. 
Symmetries and spin-multiplicities of these states are given 
in table \ref{tab1}. 
These target states were represented by valence configuration interaction 
wave functions constructed by the state averaged complete active space SCF 
(SA-CASSCF) orbitals. 
Fixed-bond approximation was employed with internuclear distances of 
3.036, 3.700 and 4.400 a$_0$ for B$_2$, S$_2$ and Si$_2$, 
respectively. 
Although we study only elastic scattering in this work, 
we included these excited target states to improve quality of 
the R-matrix calculations. 
Also, by including these excited states, we can suppress artificial 
structure coming from pseudo-resonance.  
In this study, the SA-CASSCF orbitals were obtained by calculations with 
MOLPRO suites of programs \cite{molpro}. 
The target orbitals of B$_2$, S$_2$ and Si$_2$ were constructed from the 
cc-pVTZ basis set\cite{1989JChPh..90.1007D,1993JChPh..98.1358W}. 
The radius of the R-matrix sphere was chosen to be 13 a$_0$, 
which is larger than the R-matrix sphere used in 
the electron O$_2$ calculation. We need this extended R-matrix sphere to avoid overlap 
of B$_2$, S$_2$ and Si$_2$ molecular orbitals with the R-matrix boundary. 
In order to represent the scattering electron, we included diffuse
gaussian functions up to $l$ = 4, with 13 functions for $l$ = 0, 
11 functions for $l$ = 1, 10 functions for $l$ = 2, 
8 functions for $l$ = 3, 6 functions for $l$ = 4. 
Exponents of these diffuse gaussians were taken from Faure 
et al. \cite{Fa02}. 
%
The construction of the configuration state functions (CSFs) for the electron-molecule 
system is the same as in our previous e-O$_2$ papers \cite{2006PhRvA..73e2707T,2006PhRvA..74b2706T}. 
Two different kind of ($N+1$)-electron configurations are included, where $N$ is 
a number of electrons in the target molecule. 
The first type of the ($N+1$)-electron CSFs is constructed from $N$ target molecular orbitals (MOs) plus 
one continuum orbital. 
The second type of CSFs is constructed from the $N+1$ target MOs. 
These target MOs are just the SA-CASSCF orbitals, whereas the continuum orbitals are obtained by 
orthogonalization of the diffuse gaussian functions to the target MOs \cite{Mo98}. 
Since only the continuum orbitals have overlap with the R-matrix sphere, 
the first type of CSFs mainly contributes the cross sections. 
However, the second type of CSFs is also important, as it is crucial to describe resonance. 
For reference, we show orbital set used in the e-B$_2$ calculation in 
table \ref{tab2}. The orbital sets for e-S$_2$ and e-Si$_2$ scatterings 
are very similar. 
More detailed explanation can be found in our previous paper\cite{2006PhRvA..73e2707T}.


\section{Results and discussion}

\subsection{Excitation energies}



In this section, we show excited state energies of B$_2$, S$_2$ and 
Si$_2$ molecules. Since O$_2$ energies have been shown in our previous 
paper\cite{2006PhRvA..73e2707T}, we do not discuss them here. 
In table \ref{tab3}, calculated excitation energies of B$_2$ molecule are compared 
with full configuration interaction (FCI) vertical excitation energies 
of Hald et al.\cite{2001JChPh.115..671H}.
Although they employed different basis set, aug-cc-pVDZ, and shorter internuclear distance 
of 3.005a$_0$, our CASSCF values agree reasonably well with their results. 
In table \ref{tab4}, our CASSCF energies of S$_2$ molecule are compared 
with MRD CI vertical excitation energies of Hess et al.\cite{ChemPhys.71.79} and MRCI adiabatic 
excitation energies of Kiljunen et al.\cite{2000JChPh.112.7475K}. 
In this case, our results agree well with the previous calculations for the lowest two excitations. 
For excitation energies to the three higher states, deviations become larger 
because Kiljunen et al.\cite{2000JChPh.112.7475K} studied adiabatic excitation energies 
whereas we calculated vertical excitation energies. 
In table \ref{tab5}, calculated energies of Si$_2$ molecule are compared 
with MRD CI vertical excitation energies of Peyerimhoff and Buenker \cite{ChemPhys.72.111}. 
Since they employed shorter internuclear distance of 4.3 a$_0$ compared 
to 4.4 a$_0$ of our calculation, precise comparison is difficult. 
However, our CASSCF results agree reasonably well with their results. 

\subsection{Integral cross sections}

\begin{figure}
 \includegraphics{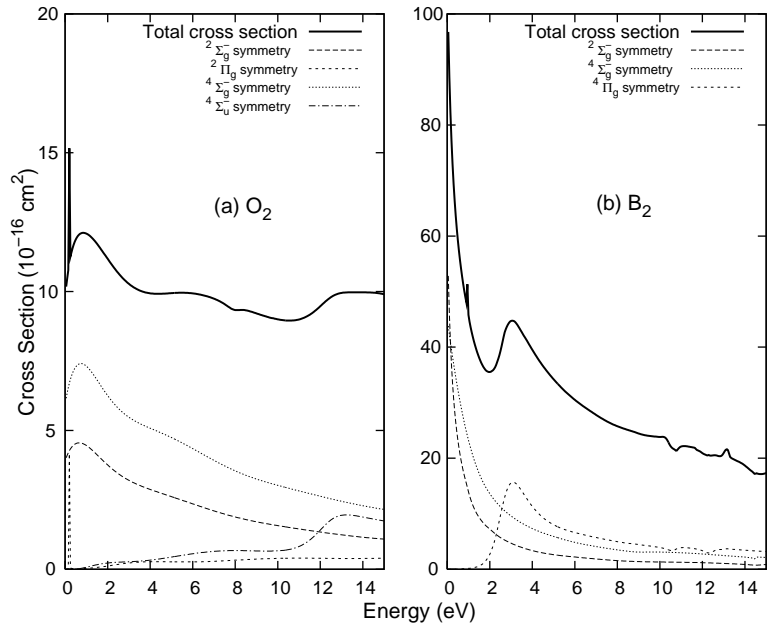}%
 \caption{\label{fig1} 
  Panel (a): the elastic integrated cross sections (ICSs) of electron scattering by O$_2$ molecules. 
  Panel (b): the elastic ICSs of electron scattering by B$_2$ molecules.
  Thick full lines represent total cross sections.  
  The partial cross sections are represented 
  by thin lines. Symmetries with minor contributions 
  are not shown in the figure.  
   }
\end{figure}

In figure \ref{fig1} (a), integral cross sections (ICSs) for elastic electron 
collision with O$_2$ molecules are shown. 
The sharp peak around 0.2 eV comes from the O$_2^-$ ${}^2 \Pi_g$ resonance. 
Also, ${}^4 \Sigma_u^-$ resonance causes a small rise of cross section around 13eV. 
The details of these e-O$_2$ ICSs were discussed in the previous paper\cite{2006PhRvA..73e2707T}, 
however, they are shown here for comparison with the ICSs of the electron B$_2$, S$_2$ and Si$_2$ 
collisions. 

In figure \ref{fig1} (b), elastic ICSs for electron B$_2$ collisions are shown. 
In this case, very large cross section is observed near zero energy,
about $10^{-14} {\rm cm}^2$, compared to $10^{-16} {\rm cm}^2$ in the e-O$_2$ collisions. 
The partial cross sections of ${}^2 \Sigma_g^{-}$ and ${}^4 \Sigma_g^{-}$ symmetries 
equally contribute to this enhancement. 
%
There is a broad peak around 3 eV, which comes from ${}^4 \Pi_g$ symmetry 
partial cross section. An analysis of configuration state functions (CSFs) suggests 
that this peak is related to the configuration $(1\sigma_g)^2(1\sigma_u)^2(2\sigma_g)^2(2\sigma_u)^2
(1\pi_u)^2(1\pi_g)^{1}$, which is the ground state B$_2$ with a scattering electron 
attached to the $1\pi$ orbital. 
By removing this $(1\pi_g)^{1}$ configuration from the R-matrix calculation, 
this peak vanishes from the ICSs. 

\begin{figure}
 \includegraphics{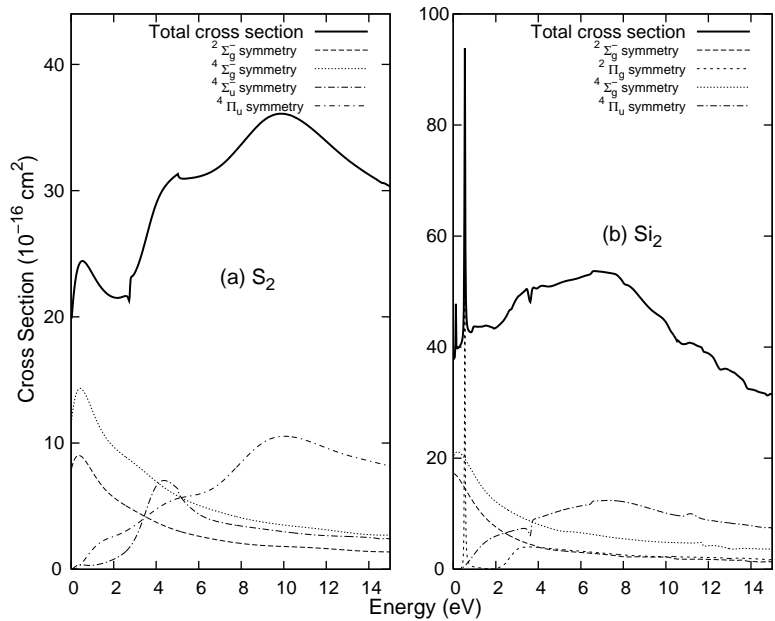}%
 \caption{\label{fig2} 
  Panel (a): the elastic ICSs of electron scattering by S$_2$ molecules. 
  Panel (b): the elastic ICSs of electron scattering by Si$_2$ molecules. 
  Other details are the same as in the figure \ref{fig1}.
  }
\end{figure}

In figure \ref{fig2} (a), ICSs for elastic electron scattering with S$_2$ molecules are shown. 
The magnitude of the ICS increases from $2.0 \times 10^{-15}$ at zero energy  
to $3.5 \times 10^{-15} {\rm cm}^2$ at 10 eV, then it decreases to 
$3.0 \times 10^{-15} {\rm cm}^2$ at 15 eV. 
Although the magnitudes are different, the profiles of the ${}^2 \Sigma_g^{-}$ and 
${}^4 \Sigma_g^{-}$ symmetry partial cross sections are very similar to those partial 
cross sections in the e-O$_2$ elastic collision. 
A broad peak is observed around 4.5 eV, which comes from the ${}^4 \Sigma_u^{-}$ 
symmetry partial cross section. The CSF analysis suggests that this peak is related 
to the configuration 
$({\rm core})^{20}(4\sigma_g)^2(4\sigma_u)^2(5\sigma_g)^2(2\pi_u)^4(2\pi_g)^2(5\sigma_u)^1$, 
and likely belongs to the S$_2^-$ ${}^4 \Sigma_u^{-}$ resonance. 
The ${}^2 \Sigma_u^{-}$ symmetry partial cross section has also a small rise 
around 6 eV (not shown in the figure), its contribution to the total ICS is small. 
Two anomalous structures are observed in the ICS, a kink at 2.7 eV and a cusp at 5 eV. 
The former kink belongs to the ${}^2 \Pi_u$ partial cross section, 
whereas The cusp at 5 eV comes from the ${}^4 \Sigma_u^{-}$ symmetry. 
We analyzed the CSFs and found that the kink at 2.7 eV is likely related to 
a resonance with configuration $({\rm core})^{20}(4\sigma_g)^2(4\sigma_u)^2(5\sigma_g)^2(2\pi_u)^3(2\pi_g)^4$, 
which is obtained from an attachment of the scattering electron to the excited 
${c}^1 \Sigma_u^-$, ${A'}^3 \Delta_u$ and ${A}^3 \Sigma_u^+$ states of S$_2$ with 
configuration $({\rm core})^{20}(4\sigma_g)^2(4\sigma_u)^2(5\sigma_g)^2(2\pi_u)^3(2\pi_g)^3$. 
The position of the cusp coincides with the S$_2$ B${}^3 \Sigma_u^{-}$ state, 
thus it is associated with opening of this excitation channel. 

In figure \ref{fig2} (b), ICSs for elastic electron scattering with Si$_2$ molecules are shown. 
The magnitude of the ICS is about $3.0 - 5.0 \times 10^{-15} {\rm cm}^2$ 
between 0 and 15 eV. 
There are two sharp peaks below 1 eV. The peak at 0.55 eV is from the ${}^2 \Pi_g$ 
symmetry partial cross sections and the other peak at 0.12 eV is from the ${}^4 \Pi_g$ symmetry. 
We checked the CSFs of the ${}^2 \Pi_g$ and  ${}^4 \Pi_g$ symmetry calculations and found that 
the configuration $({\rm core})^{20}(4\sigma_g)^2(4\sigma_u)^2(5\sigma_g)^2(2\pi_u)^2(2\pi_g)^1$
has dominant contribution to these resonances. 

\subsection{Polarization fractions}

\begin{figure}
 \includegraphics{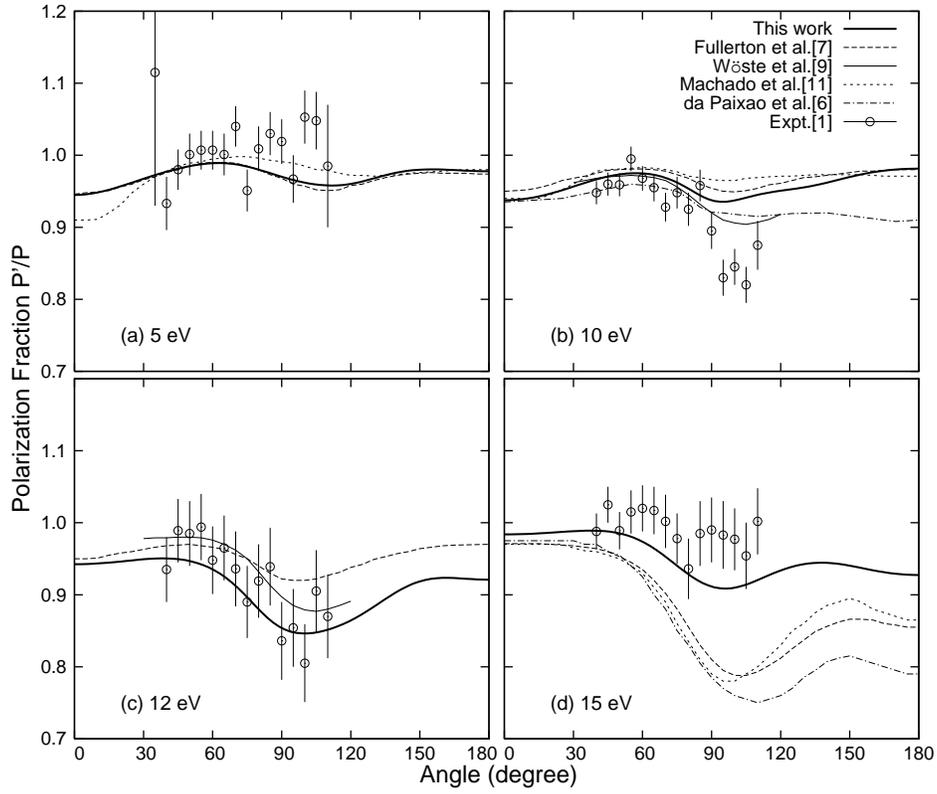}%
 \caption{\label{fig3} 
  Polarization fractions of electron O$_2$ elastic scattering. 
  Panel (a): 5eV, (b) 10eV, (c) 12eV, (d) 15eV. 
  Our results are shown as thick full lines. 
  Experimental results of Hegemann et al.\cite{1993JPhB...26.4607H} 
  are shown as open circles with error bars and  
  theoretical PFs of Fullerton et al.\cite{1994JPhB...27..185F}, 
  Machado et al.\cite{1999PhRvA..60.1199M}, da Paixao et al.\cite{1992PhRvL..68.1698D}, 
  and W\"oste et al.\cite{1996JPhB...29.2553W} are shown as thin lines. 
  }
\end{figure}

In figure \ref{fig3}, calculated polarization fractions (PFs) for 
elastic electron O$_2$ collisions are shown for scattering  
energies of 5, 10, 12 and 15 eV with the previous theoretical results 
of Fullerton et al.\cite{1994JPhB...27..185F}, 
Machado et al.\cite{1999PhRvA..60.1199M}, da Paix\~ao et al.\cite{1992PhRvL..68.1698D} and 
W\"oste et al.\cite{1996JPhB...29.2553W}.  
These theoretical results are also compared with the experimental values of 
Hegemann et al.\cite{1993JPhB...26.4607H} in the figure. 
As we can see from eq.\ref{eq1}, deviation of PF from unity is a measure of spin-exchange. 
Our e-O$_2$ PFs are close to 1 at all scattering energies, indicating the degree of spin-exchange 
is relatively small.  
For 5 eV, our results are very similar to the previous R-matrix 
results of Fullerton et al.\cite{1994JPhB...27..185F}. 
The results of Machado et al.\cite{1999PhRvA..60.1199M} are also similar, but smaller at low angles below 
30 degrees. 
Our PFs at 10eV are slightly smaller than the results of Fullerton et al.\cite{1994JPhB...27..185F} 
and Machado et al.\cite{1999PhRvA..60.1199M}. 
The PFs of da Paixao et al.\cite{1992PhRvL..68.1698D} at 10 eV are smaller than our results in all 
angles, especially 120-180 degrees.  
Our calculation cannot reproduce the drop of experimental PFs at 10 eV at 100 degrees.  
The PFs of W\"oste et al.\cite{1996JPhB...29.2553W} have a minimum at this position and their value at 100 degree 
is the closest to the experimental result, although there is still some deviation in magnitude.  
At collision energy of 12 eV, the results of our calculation, W\"oste et al.\cite{1996JPhB...29.2553W} and 
Fullerton et al.\cite{1994JPhB...27..185F} have similar angular behaviour, though our results are smaller 
than the others at all angles. 
All of these three theoretical results agree reasonably well with the experimental results. 
For 15eV, our PFs are larger than the results of the other theoretical 
calculations in all scattering angles and are closer to the experimental results. 
The PFs of Fullerton et al.\cite{1994JPhB...27..185F} and Machado et al.\cite{1999PhRvA..60.1199M} are 
very similar in shape and magnitude, whereas the results of da Paixao et al.\cite{1992PhRvL..68.1698D} are 
slightly smaller at higher angles above 110 degrees. 
The deviation of our PFs from the previous theoretical results is the largest around 90-110 degrees, 
where there is a dip in the profile. Although the magnitude of the PFs are different, 
the shape of the our PF profiles itself is similar to the previous calculations. 

\begin{figure}
 \includegraphics{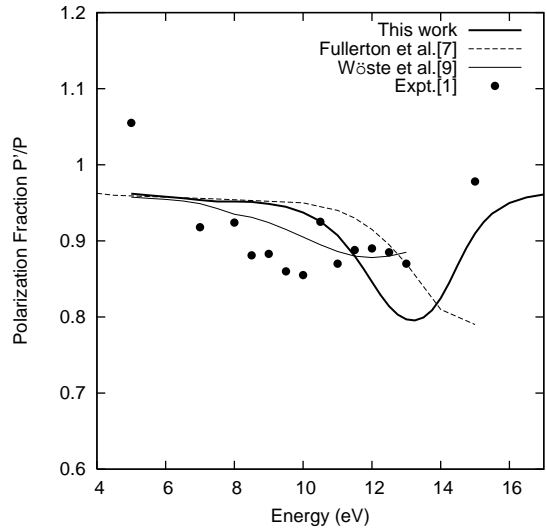}%
 \caption{\label{fig4}
  Polarization fractions of electron O$_2$ elastic scattering, 
  as a function of scattering energy at an scattering angle of 
  100 degrees.  Other details are the same as in the figure \ref{fig3}.
  }
\end{figure}

In figure \ref{fig4}, the PFs for elastic electron O$_2$ collisions are shown 
as a function of energy at a scattering angle of 100 degrees. 
Our result has a minimum at 13 eV, however, it is located at 15 eV and 12 eV in the result of 
Fullerton et al.\cite{1994JPhB...27..185F} and W\"oste et al.\cite{1996JPhB...29.2553W}, respectively. 
The magnitude of the PF at the minimum is larger in W\"oste et al.\cite{1996JPhB...29.2553W} than in our calculation 
and Fullerton et al.\cite{1994JPhB...27..185F}. 

\begin{figure}
 \includegraphics{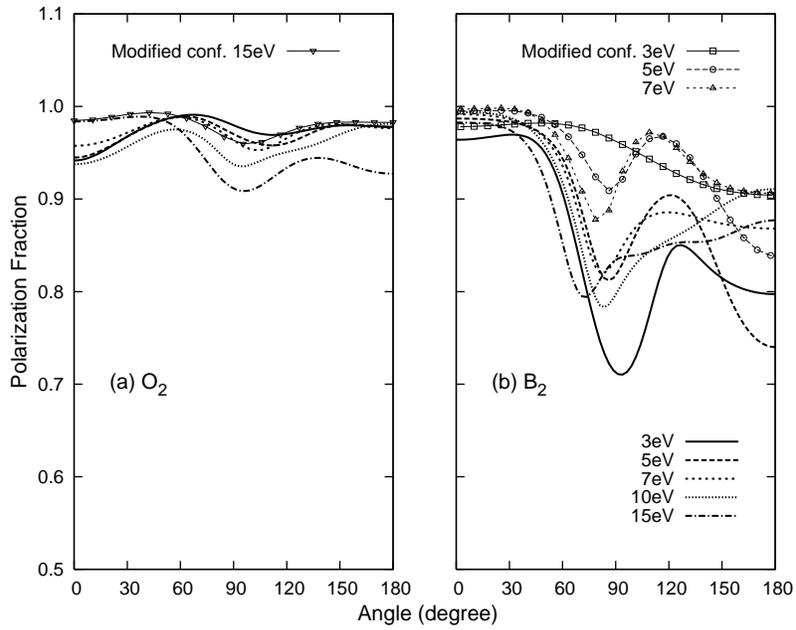}%
 \caption{\label{fig5} 
  Panel (a): polarization fractions (PFs) of electron O$_2$ elastic scattering. 
  Panel (b): PFs of electron B$_2$ elastic scattering. 
  Calculated PFs are shown as thick lines. 
  Thin lines with symbols represent PFs without the effect of resonances, obtained 
  by the R-matrix calculation with modified configurations (see text for details). 
  }
\end{figure}

For comparison of the e-O$_2$ PFs with the PFs of electron B$_2$, S$_2$ and Si$_2$ collisions 
in the following figures,  
the PFs of elastic electron O$_2$ collisions are again shown in the figure \ref{fig5} (a) 
for collision energies of 3, 5, 7, 10 and 15 eV. 
The depolarization, i.e., deviation of PF from 1, is only prominent at 10 and 15 eV where 
the ${}^4 \Sigma_u^-$ resonance exists as shown in fig.\ref{fig1} (a). 
In order to check the relation of the ${}^4 \Sigma_u^-$ resonance and the PFs 
at 15 eV, we have carried out the R-matrix calculation with modified configurations, 
removing the $(1\sigma_g)^2(1\sigma_u)^2(2\sigma_g)^2(2\sigma_u)^2(3\sigma_g)^2(1\pi_u)^4(1\pi_g)^{2}(3\sigma_u)^{1}$ 
configuration from the original calculation. 
By this procedure we can suppress the effect of the resonance. 
The results in the fig. \ref{fig5} (a) indicates that 
the PFs become very close to 1 by removing the configuration of the ${}^4 \Sigma_u^-$ resonance.  

In figure \ref{fig5} (b), calculated PFs for elastic electron B$_2$ collisions 
are shown for scattering energies of 3, 5, 7, 10 and 15 eV. 
For most of the scattering energies and angles, the depolarization in electron B$_2$ collision is 
larger than that in electron O$_2$ collisions. 
Between 80 and 180 degrees, the magnitude of the PFs are about 0.8-0.9 at all energies.  
In contrast, the e-O$_2$ PFs are larger than 0.9.   
The e-B$_2$ PFs show large depolarization effect at scattering energies of 3 and 5 eV, 
which are close to the ${}^4 \Pi_g$ resonance. 
In order to understand the origin of the large depolarizations at 3 and 5 eV, 
we excluded the effect of the B$_2^-$ ${}^4 \Pi_g$ resonance around 3.5 eV and 
re-calculated the e-B$_2$ PFs. 
Specifically, we removed $(1\sigma_g)^2(1\sigma_u)^2(2\sigma_g)^2(2\sigma_u)^2
(1\pi_u)^2(1\pi_g)^{1}$ configuration from the R-matrix calculation 
and erased the ${}^4 \Pi_g$ resonance. 
As shown in the fig. \ref{fig5} (b), the effect of the resonance on the PFs is evident. 
With the resonance effect, the lowest value of the PF at 3 eV is about 0.7 at 90 degrees, 
but it becomes about 0.95 without the resonance contribution. 
Also, the depolarizations at 5 and 7 eV become less pronounced when we remove 
the effect of the resonance. 

\begin{figure}
 \includegraphics{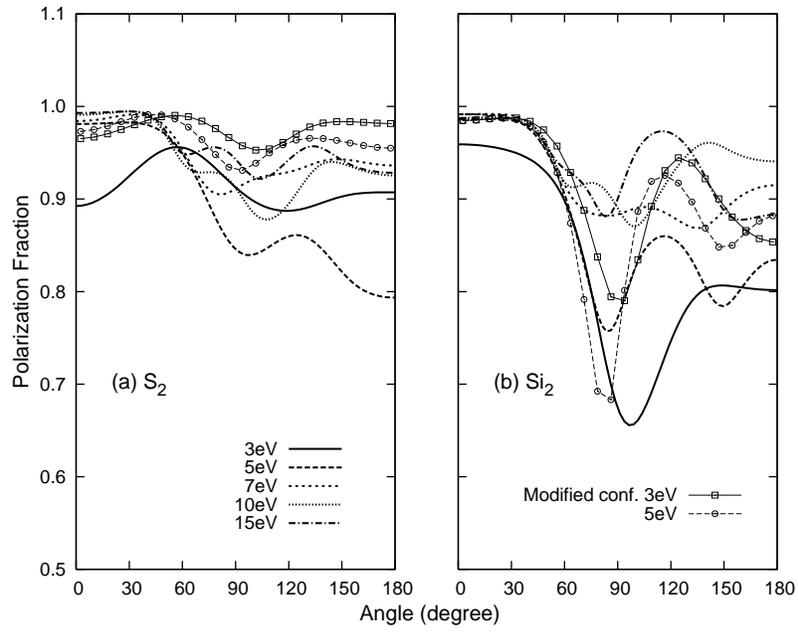}%
 \caption{\label{fig6} 
  Panel (a): polarization fractions (PFs) of electron S$_2$ elastic scattering. 
  Panel (b): PFs of electron Si$_2$ elastic scattering. 
  Other details are the same as in the figure \ref{fig5}.
  }
\end{figure}

The calculated PFs for elastic electron S$_2$ collisions are shown in figure \ref{fig6} (a). 
In general, the degree of depolarization is smaller than the e-B$_2$ case, 
but is larger than the e-O$_2$ case. The profiles of the PFs at 7, 10 and 15 eV look similar 
to each other. However, the PFs at 3 and 5 eV behave differently. 
The degree of depolarization is larger at forward angles for 3 eV case, however, 
it is larger at backward angles at 5 eV. 
To understand the effect of resonance on electron S$_2$ PFs, we removed the 
$({\rm core})^{20}(4\sigma_g)^2(4\sigma_u)^2(5\sigma_g)^2(2\pi_u)^4(2\pi_g)^2(5\sigma_u)^1$ 
configuration and erased the ${}^4 \Sigma_u^{-}$ resonance at 4.5eV, 
then re-calculated the PFs. The results are shown in the same figure. 
As in the case of electron B$_2$ PFs, the degree of depolarization becomes smaller 
when we removed the resonance effect. 

The calculated PFs for elastic electron Si$_2$ collisions are shown in figure \ref{fig6} (b). 
In this case, relatively large depolarization is observed for 3 eV at 100 degrees. 
The depolarization becomes smaller as the collision energy increases,  
however, some degree of depolarization remains near 80 and 180 degrees. 
The effect of resonance on the electron Si$_2$ PFs were examined by removing the 
$({\rm core})^{20}(4\sigma_g)^2(4\sigma_u)^2(5\sigma_g)^2(2\pi_u)^2(2\pi_g)^1$ 
configuration, which is responsible for the ${}^2 \Pi_g$ and  ${}^4 \Pi_g$ resonances 
at 0.12 and 0.55 eV, respectively. By removing this configuration, these two sharp peaks 
in the ICSs disappear. 
The PFs without the resonances are shown in the fig. \ref{fig6} (b). 
For 3 eV case, the depolarization becomes smaller at all angles. However, the decrease of 
depolarization is not so large compared to the cases of e-B$_2$ and e-S$_2$ collisions. 
For 5 eV case, the depolarization at 85 degrees becomes larger, though it becomes smaller at 
backward angles. Thus, in this case, the association of the resonances with the depolarization is 
not so straightforward as in the e-O$_2$, B$_2$ and S$_2$ cases. 

\subsection{Discussion}

As we show in the figures \ref{fig5} and \ref{fig6}, existence of resonance and 
behaviour of PF is closely related each other. When resonance exists at some energy, 
relatively large depolarization is observed compared to the other energies. 
Also, when the resonance is artificially removed by deleting specific configuration 
in the R-matrix calculation, depolarization becomes smaller in general.  
In case of electron Si$_2$ collisions, this trend is partly broken at 5 eV around 
80 degrees, however, depolarization generally becomes smaller at the other region 
after removing the resonance effect. 
The association of resonance and PF has been discussed in the previous theoretical and 
experimental papers \cite{1993JPhB...26.4607H, 1994JPhB...27..185F, 1996JPhB...29.2553W}, 
and we have confirmed this association more clearly by explicitly studying the effect of 
resonance on the PF of four different electron-molecule systems. 

Even if collision occurs away from the resonance energy, some degree of depolarization 
is observed in all cases of e-O$_2$, B$_2$, S$_2$ and Si$_2$ collisions. 
In e-B$_2$ case, depolarization is relatively large even outside of the resonance energy region. 
In contrast, the PFs in e-O$_2$ collisions are very close to 1 when collision energy is distant from 
the resonance energy. The degree of depolarization in e-S$_2$ and Si$_2$ collisions is intermediate 
between e-B$_2$ and e-O$_2$ depolarizations. 
It is unclear why different degree of depolarization is seen in these 
four electron-molecule collisions when collision energy is distant from resonance. 
The extent of molecular orbitals may be related to this difference, 
as discussed by Sartori et al.\cite{PhysRevA.55.3243} 
on the electron H$_2$ superelastic collisions. 


For the electron O$_2$ elastic scattering, the previous theoretical and experimental 
PFs are available for comparison with our results. Our low energy PFs at 5 and 10 eV 
are similar to the other theoretical results. However, the PFs at 12eV 
are smaller than the previous results at all angles, and our PFs at 15eV are 
much closer to unity than the other theoretical results as shown in fig.\ref{fig3}. 
The reason of these deviations can be attributed to the shift of O$_2^-$ ${}^4 \Sigma_u^-$ 
resonance position. In our calculation, the position of the ${}^4 \Sigma_u^-$ 
resonance peak is located around 13.0 eV, whereas it is around 14.1 eV in the previous 
R-matrix calculation. 
As discussed in W\"oste et al.\cite{1996JPhB...29.2553W}, the position of 
the ${}^4 \Sigma_u^-$ resonance is sensitive to the internuclear distance. 
In this work, the internuclear length is fixed to be 2.3 a0 and it is the same as 
in the calculation of Fullerton et al.\cite{1994JPhB...27..185F}. 
So the choice of basis set, molecular orbitals and number of target states is important 
to the difference in the position of the resonance.       
Probably the position of resonance is stabilized by inclusion 
of more target states in the present R-matrix calculations compared 
to the previous calculations of Fullerton et al.\cite{1994JPhB...27..185F}, 
Machado et al.\cite{1999PhRvA..60.1199M} and da Paix\~ao et al.\cite{1992PhRvL..68.1698D}.   

In figure \ref{fig1} (b), sharp increase of cross section is observed in electron 
B$_2$ elastic scattering near zero energy. 
This increase of cross section is similar to the case of electron polar-molecule collision, 
although B$_2$ molecule has no dipole moment. 
The cross section of electron CO$_2$ elastic collision also has similar sharp 
increase near zero energy, and several experimental and theoretical works have been performed to 
understand this behaviour. Morrison analyzed this problem and suggested that this behaviour 
is related to the existence of a virtual state \cite{PhysRevA.25.1445}. 
Morgan has shown that the correlation and 
polarization effect is important for this sharp peak \cite{1998PhRvL..80.1873M}. 
Similar mechanism may exist for the electron B$_2$ elastic collisions. 
%
%

For the electron O$_2$ elastic collisions, we put larger number of target electronic 
states and better quality molecular orbitals in the R-matrix calculations, and 
obtained improved results around ${}^4 \Sigma_u^-$ resonance region compared to 
the previous fixed-bond calculations. 
However, the results of W\"oste et al.\cite{1996JPhB...29.2553W} agree better with the experimental PFs 
at 10 eV. They achieved this good agreement by vibrational averaging of the T-matrices to 
include the effect of the nuclear motion. 
By extending the present R-matrix calculation to include the vibrational 
effect using vibrational averaging procedure or the non-adiabatic R-matrix method, 
we may obtain better agreement with the experimental PFs.  
Also, inclusion of nuclear motion effect may improve the quality of the calculations 
on electron B$_2$, S$_2$ and Si$_2$ elastic collisions. 


\section{summary}

We have calculated the polarization fractions (PFs) on low-energy elastic 
collisions of spin-polarized electrons with open-shell molecules, O$_2$, B$_2$, 
S$_2$ and Si$_2$, all of them having ${}^3 \Sigma_g^-$ symmetry in their ground states. 
As in our previous works, we employed the fixed-bond R-matrix method based 
on state-averaged complete active space SCF orbitals. 
Our PFs for electron O$_2$ collisions agree better with the previous 
experimental result, especially around the ${}^4 \Sigma_u^-$ resonance, compared to 
the previous theoretical calculations. 
Larger spin-exchange effect is observed in the electron B$_2$ and Si$_2$ collisions 
than in the e-O$_2$ collisions. In e-S$_2$ collisions, degree of depolarization is 
similar to the e-O$_2$ collisions. 
In all four electron-molecule collisions, the PFs deviate larger from 1 near resonances. 
This association of resonance and PF was explicitly confirmed by the R-matrix calculations 
removing configurations responsible for the resonance. 


%


\begin{thebibliography}{29}
\expandafter\ifx\csname natexlab\endcsname\relax\def\natexlab#1{#1}\fi
\expandafter\ifx\csname bibnamefont\endcsname\relax
  \def\bibnamefont#1{#1}\fi
\expandafter\ifx\csname bibfnamefont\endcsname\relax
  \def\bibfnamefont#1{#1}\fi
\expandafter\ifx\csname citenamefont\endcsname\relax
  \def\citenamefont#1{#1}\fi
\expandafter\ifx\csname url\endcsname\relax
  \def\url#1{\texttt{#1}}\fi
\expandafter\ifx\csname urlprefix\endcsname\relax\def\urlprefix{URL }\fi
\providecommand{\bibinfo}[2]{#2}
\providecommand{\eprint}[2][]{\url{#2}}

\bibitem[{\citenamefont{{Hegemann} et~al.}(1993)\citenamefont{{Hegemann},
  {Schroll}, and {Hanne}}}]{1993JPhB...26.4607H}
\bibinfo{author}{\bibfnamefont{T.}~\bibnamefont{{Hegemann}}},
  \bibinfo{author}{\bibfnamefont{S.}~\bibnamefont{{Schroll}}},
  \bibnamefont{and} \bibinfo{author}{\bibfnamefont{G.~F.}
  \bibnamefont{{Hanne}}}, \bibinfo{journal}{J. Phys. B}
  \textbf{\bibinfo{volume}{26}}, \bibinfo{pages}{4607} (\bibinfo{year}{1993}).

\bibitem[{\citenamefont{{Huang}}(1987)}]{1987ChJPh..25..156H}
\bibinfo{author}{\bibfnamefont{K.-N.} \bibnamefont{{Huang}}},
  \bibinfo{journal}{Chinese Journal of Physics} \textbf{\bibinfo{volume}{25}},
  \bibinfo{pages}{156} (\bibinfo{year}{1987}).

\bibitem[{\citenamefont{{Bartschat} and {Madison}}(1988)}]{1988JPhB...21.2621B}
\bibinfo{author}{\bibfnamefont{K.}~\bibnamefont{{Bartschat}}} \bibnamefont{and}
  \bibinfo{author}{\bibfnamefont{D.~H.} \bibnamefont{{Madison}}},
  \bibinfo{journal}{J. Phys. B} \textbf{\bibinfo{volume}{21}},
  \bibinfo{pages}{2621} (\bibinfo{year}{1988}).

\bibitem[{\citenamefont{Hegemann et~al.}(1991)\citenamefont{Hegemann,
  Oberste-Vorth, Vogts, and Hanne}}]{PhysRevLett.66.2968}
\bibinfo{author}{\bibfnamefont{T.}~\bibnamefont{Hegemann}},
  \bibinfo{author}{\bibfnamefont{M.}~\bibnamefont{Oberste-Vorth}},
  \bibinfo{author}{\bibfnamefont{R.}~\bibnamefont{Vogts}}, \bibnamefont{and}
  \bibinfo{author}{\bibfnamefont{G.~F.} \bibnamefont{Hanne}},
  \bibinfo{journal}{Phys. Rev. Lett.} \textbf{\bibinfo{volume}{66}},
  \bibinfo{pages}{2968} (\bibinfo{year}{1991}).

\bibitem[{\citenamefont{{Ratliff} et~al.}(1989)\citenamefont{{Ratliff},
  {Rutherford}, {Dunning}, and {Walters}}}]{1989PhRvA..39.5584R}
\bibinfo{author}{\bibfnamefont{J.~M.} \bibnamefont{{Ratliff}}},
  \bibinfo{author}{\bibfnamefont{G.~H.} \bibnamefont{{Rutherford}}},
  \bibinfo{author}{\bibfnamefont{F.~B.} \bibnamefont{{Dunning}}},
  \bibnamefont{and} \bibinfo{author}{\bibfnamefont{G.~K.}
  \bibnamefont{{Walters}}}, \bibinfo{journal}{\pra}
  \textbf{\bibinfo{volume}{39}}, \bibinfo{pages}{5584} (\bibinfo{year}{1989}).

\bibitem[{\citenamefont{{da Paix\~ao} et~al.}(1992)\citenamefont{{da Paix\~ao},
  {Lima}, and {McKoy}}}]{1992PhRvL..68.1698D}
\bibinfo{author}{\bibfnamefont{F.~J.} \bibnamefont{{da Paix\~ao}}},
  \bibinfo{author}{\bibfnamefont{M.~A.~P.} \bibnamefont{{Lima}}},
  \bibnamefont{and} \bibinfo{author}{\bibfnamefont{V.}~\bibnamefont{{McKoy}}},
  \bibinfo{journal}{Phys. Rev. Lett.} \textbf{\bibinfo{volume}{68}},
  \bibinfo{pages}{1698} (\bibinfo{year}{1992}).

\bibitem[{\citenamefont{{Fullerton} et~al.}(1994)\citenamefont{{Fullerton},
  {W{\"o}ste}, {Thompson}, {Blum}, and {Noble}}}]{1994JPhB...27..185F}
\bibinfo{author}{\bibfnamefont{C.~M.} \bibnamefont{{Fullerton}}},
  \bibinfo{author}{\bibfnamefont{G.}~\bibnamefont{{W{\"o}ste}}},
  \bibinfo{author}{\bibfnamefont{D.~G.} \bibnamefont{{Thompson}}},
  \bibinfo{author}{\bibfnamefont{K.}~\bibnamefont{{Blum}}}, \bibnamefont{and}
  \bibinfo{author}{\bibfnamefont{C.~J.} \bibnamefont{{Noble}}},
  \bibinfo{journal}{J. Phys. B} \textbf{\bibinfo{volume}{27}},
  \bibinfo{pages}{185} (\bibinfo{year}{1994}).

\bibitem[{\citenamefont{{Nordbeck} et~al.}(1994)\citenamefont{{Nordbeck},
  {Fullerton}, {W{\"o}ste}, {Thompson}, and {Blum}}}]{1994JPhB...27.5375N}
\bibinfo{author}{\bibfnamefont{R.-P.} \bibnamefont{{Nordbeck}}},
  \bibinfo{author}{\bibfnamefont{C.~M.} \bibnamefont{{Fullerton}}},
  \bibinfo{author}{\bibfnamefont{G.}~\bibnamefont{{W{\"o}ste}}},
  \bibinfo{author}{\bibfnamefont{D.~G.} \bibnamefont{{Thompson}}},
  \bibnamefont{and} \bibinfo{author}{\bibfnamefont{K.}~\bibnamefont{{Blum}}},
  \bibinfo{journal}{J. Phys. B} \textbf{\bibinfo{volume}{27}},
  \bibinfo{pages}{5375} (\bibinfo{year}{1994}).

\bibitem[{\citenamefont{{W{\"o}ste} et~al.}(1996)\citenamefont{{W{\"o}ste},
  {Higgins}, {Duddy}, {Fullerton}, and {Thompson}}}]{1996JPhB...29.2553W}
\bibinfo{author}{\bibfnamefont{G.}~\bibnamefont{{W{\"o}ste}}},
  \bibinfo{author}{\bibfnamefont{K.}~\bibnamefont{{Higgins}}},
  \bibinfo{author}{\bibfnamefont{P.}~\bibnamefont{{Duddy}}},
  \bibinfo{author}{\bibfnamefont{C.~M.} \bibnamefont{{Fullerton}}},
  \bibnamefont{and} \bibinfo{author}{\bibfnamefont{D.~G.}
  \bibnamefont{{Thompson}}}, \bibinfo{journal}{J. Phys. B}
  \textbf{\bibinfo{volume}{29}}, \bibinfo{pages}{2553} (\bibinfo{year}{1996}).

\bibitem[{\citenamefont{Noble and Burke}(1992)}]{No92}
\bibinfo{author}{\bibfnamefont{C.~J.} \bibnamefont{Noble}} \bibnamefont{and}
  \bibinfo{author}{\bibfnamefont{P.~G.} \bibnamefont{Burke}},
  \bibinfo{journal}{Phys. Rev. Lett.} \textbf{\bibinfo{volume}{68}},
  \bibinfo{pages}{2011} (\bibinfo{year}{1992}).

\bibitem[{\citenamefont{{Machado} et~al.}(1999)\citenamefont{{Machado},
  {Ribeiro}, {Lee}, {Fujimoto}, and {Brescansin}}}]{1999PhRvA..60.1199M}
\bibinfo{author}{\bibfnamefont{L.~E.} \bibnamefont{{Machado}}},
  \bibinfo{author}{\bibfnamefont{E.~M.~S.} \bibnamefont{{Ribeiro}}},
  \bibinfo{author}{\bibfnamefont{M.-T.} \bibnamefont{{Lee}}},
  \bibinfo{author}{\bibfnamefont{M.~M.} \bibnamefont{{Fujimoto}}},
  \bibnamefont{and} \bibinfo{author}{\bibfnamefont{L.~M.}
  \bibnamefont{{Brescansin}}}, \bibinfo{journal}{\pra}
  \textbf{\bibinfo{volume}{60}}, \bibinfo{pages}{1199} (\bibinfo{year}{1999}).

\bibitem[{\citenamefont{da~Paix\~ao et~al.}(1996)\citenamefont{da~Paix\~ao,
  Lima, and McKoy}}]{PhysRevA.53.1400}
\bibinfo{author}{\bibfnamefont{F.~J.} \bibnamefont{da~Paix\~ao}},
  \bibinfo{author}{\bibfnamefont{M.~A.~P.} \bibnamefont{Lima}},
  \bibnamefont{and} \bibinfo{author}{\bibfnamefont{V.}~\bibnamefont{McKoy}},
  \bibinfo{journal}{Phys. Rev. A} \textbf{\bibinfo{volume}{53}},
  \bibinfo{pages}{1400} (\bibinfo{year}{1996}).

\bibitem[{\citenamefont{Sartori et~al.}(1997)\citenamefont{Sartori,
  da~Paix\~ao, and Lima}}]{PhysRevA.55.3243}
\bibinfo{author}{\bibfnamefont{C.~S.} \bibnamefont{Sartori}},
  \bibinfo{author}{\bibfnamefont{F.~J.} \bibnamefont{da~Paix\~ao}},
  \bibnamefont{and} \bibinfo{author}{\bibfnamefont{M.~A.~P.}
  \bibnamefont{Lima}}, \bibinfo{journal}{Phys. Rev. A}
  \textbf{\bibinfo{volume}{55}}, \bibinfo{pages}{3243} (\bibinfo{year}{1997}).

\bibitem[{\citenamefont{{Fujimoto} et~al.}(2006)\citenamefont{{Fujimoto},
  {Michelin}, {Iga}, and {Lee}}}]{2006PhRvA..73a2714F}
\bibinfo{author}{\bibfnamefont{M.~M.} \bibnamefont{{Fujimoto}}},
  \bibinfo{author}{\bibfnamefont{S.~E.} \bibnamefont{{Michelin}}},
  \bibinfo{author}{\bibfnamefont{I.}~\bibnamefont{{Iga}}}, \bibnamefont{and}
  \bibinfo{author}{\bibfnamefont{M.-T.} \bibnamefont{{Lee}}},
  \bibinfo{journal}{\pra} \textbf{\bibinfo{volume}{73}},
  \bibinfo{pages}{012714} (\bibinfo{year}{2006}).

\bibitem[{\citenamefont{{Tashiro}
  et~al.}(2006{\natexlab{a}})\citenamefont{{Tashiro}, {Morokuma}, and
  {Tennyson}}}]{2006PhRvA..73e2707T}
\bibinfo{author}{\bibfnamefont{M.}~\bibnamefont{{Tashiro}}},
  \bibinfo{author}{\bibfnamefont{K.}~\bibnamefont{{Morokuma}}},
  \bibnamefont{and}
  \bibinfo{author}{\bibfnamefont{J.}~\bibnamefont{{Tennyson}}},
  \bibinfo{journal}{\pra} \textbf{\bibinfo{volume}{73}},
  \bibinfo{pages}{052707} (\bibinfo{year}{2006}{\natexlab{a}}).

\bibitem[{\citenamefont{{Tashiro}
  et~al.}(2006{\natexlab{b}})\citenamefont{{Tashiro}, {Morokuma}, and
  {Tennyson}}}]{2006PhRvA..74b2706T}
\bibinfo{author}{\bibfnamefont{M.}~\bibnamefont{{Tashiro}}},
  \bibinfo{author}{\bibfnamefont{K.}~\bibnamefont{{Morokuma}}},
  \bibnamefont{and}
  \bibinfo{author}{\bibfnamefont{J.}~\bibnamefont{{Tennyson}}},
  \bibinfo{journal}{\pra} \textbf{\bibinfo{volume}{74}},
  \bibinfo{pages}{022706} (\bibinfo{year}{2006}{\natexlab{b}}).

\bibitem[{\citenamefont{Morgan et~al.}(1998)\citenamefont{Morgan, Tennyson, and
  Gillan}}]{Mo98}
\bibinfo{author}{\bibfnamefont{L.~A.} \bibnamefont{Morgan}},
  \bibinfo{author}{\bibfnamefont{J.}~\bibnamefont{Tennyson}}, \bibnamefont{and}
  \bibinfo{author}{\bibfnamefont{C.~J.} \bibnamefont{Gillan}},
  \bibinfo{journal}{Comput. Phys. Commun.} \textbf{\bibinfo{volume}{114}},
  \bibinfo{pages}{120} (\bibinfo{year}{1998}).

\bibitem[{\citenamefont{Burke and Tennyson}(2005)}]{Bu05}
\bibinfo{author}{\bibfnamefont{P.~G.} \bibnamefont{Burke}} \bibnamefont{and}
  \bibinfo{author}{\bibfnamefont{J.}~\bibnamefont{Tennyson}},
  \bibinfo{journal}{Mol. Phys.} \textbf{\bibinfo{volume}{103}},
  \bibinfo{pages}{2537} (\bibinfo{year}{2005}).

\bibitem[{\citenamefont{Gorfinkiel et~al.}(2005)\citenamefont{Gorfinkiel,
  Faure, Taioli, Piccarreta, Halmova, and Tennyson}}]{Go05}
\bibinfo{author}{\bibfnamefont{J.~D.} \bibnamefont{Gorfinkiel}},
  \bibinfo{author}{\bibfnamefont{A.}~\bibnamefont{Faure}},
  \bibinfo{author}{\bibfnamefont{S.}~\bibnamefont{Taioli}},
  \bibinfo{author}{\bibfnamefont{C.}~\bibnamefont{Piccarreta}},
  \bibinfo{author}{\bibfnamefont{G.}~\bibnamefont{Halmova}}, \bibnamefont{and}
  \bibinfo{author}{\bibfnamefont{J.}~\bibnamefont{Tennyson}},
  \bibinfo{journal}{Eur. Phys. J. D} \textbf{\bibinfo{volume}{35}},
  \bibinfo{pages}{231} (\bibinfo{year}{2005}).

\bibitem[{\citenamefont{Werner et~al.}()\citenamefont{Werner, Knowles, Lindh,
  {Sch\"{u}tz} et~al.}}]{molpro}
\bibinfo{author}{\bibfnamefont{H.-J.} \bibnamefont{Werner}},
  \bibinfo{author}{\bibfnamefont{P.~J.} \bibnamefont{Knowles}},
  \bibinfo{author}{\bibfnamefont{R.}~\bibnamefont{Lindh}},
  \bibinfo{author}{\bibfnamefont{M.}~\bibnamefont{{Sch\"{u}tz}}},
  \bibnamefont{et~al.}, \emph{\bibinfo{title}{Molpro version 2002.6, a package
  of ab initio programs}}.

\bibitem[{\citenamefont{{Dunning}}(1989)}]{1989JChPh..90.1007D}
\bibinfo{author}{\bibfnamefont{T.~H.} \bibnamefont{{Dunning}},
  \bibfnamefont{Jr.}}, \bibinfo{journal}{\jcp} \textbf{\bibinfo{volume}{90}},
  \bibinfo{pages}{1007} (\bibinfo{year}{1989}).

\bibitem[{\citenamefont{{Woon} and {Dunning}}(1993)}]{1993JChPh..98.1358W}
\bibinfo{author}{\bibfnamefont{D.~E.} \bibnamefont{{Woon}}} \bibnamefont{and}
  \bibinfo{author}{\bibfnamefont{T.~H.} \bibnamefont{{Dunning}},
  \bibfnamefont{Jr.}}, \bibinfo{journal}{\jcp} \textbf{\bibinfo{volume}{98}},
  \bibinfo{pages}{1358} (\bibinfo{year}{1993}).

\bibitem[{\citenamefont{Faure et~al.}(2002)\citenamefont{Faure, Gorfinkiel,
  Morgan, and Tennyson}}]{Fa02}
\bibinfo{author}{\bibfnamefont{A.}~\bibnamefont{Faure}},
  \bibinfo{author}{\bibfnamefont{J.~D.} \bibnamefont{Gorfinkiel}},
  \bibinfo{author}{\bibfnamefont{L.~A.} \bibnamefont{Morgan}},
  \bibnamefont{and} \bibinfo{author}{\bibfnamefont{J.}~\bibnamefont{Tennyson}},
  \bibinfo{journal}{Comput. Phys. Commun.} \textbf{\bibinfo{volume}{144}},
  \bibinfo{pages}{224} (\bibinfo{year}{2002}).

\bibitem[{\citenamefont{{Hald} et~al.}(2001)\citenamefont{{Hald},
  {J{\o}rgensen}, {Olsen}, and {Jaszu{\'n}ski}}}]{2001JChPh.115..671H}
\bibinfo{author}{\bibfnamefont{K.}~\bibnamefont{{Hald}}},
  \bibinfo{author}{\bibfnamefont{P.}~\bibnamefont{{J{\o}rgensen}}},
  \bibinfo{author}{\bibfnamefont{J.}~\bibnamefont{{Olsen}}}, \bibnamefont{and}
  \bibinfo{author}{\bibfnamefont{M.}~\bibnamefont{{Jaszu{\'n}ski}}},
  \bibinfo{journal}{\jcp} \textbf{\bibinfo{volume}{115}}, \bibinfo{pages}{671}
  (\bibinfo{year}{2001}).

\bibitem[{\citenamefont{Hess et~al.}(1982)\citenamefont{Hess, Buenker, Marian,
  and Peyerimhoff}}]{ChemPhys.71.79}
\bibinfo{author}{\bibfnamefont{B.}~\bibnamefont{Hess}},
  \bibinfo{author}{\bibfnamefont{R.~J.} \bibnamefont{Buenker}},
  \bibinfo{author}{\bibfnamefont{C.~M.} \bibnamefont{Marian}},
  \bibnamefont{and} \bibinfo{author}{\bibfnamefont{S.~D.}
  \bibnamefont{Peyerimhoff}}, \bibinfo{journal}{Chem. Phys.}
  \textbf{\bibinfo{volume}{71}}, \bibinfo{pages}{79} (\bibinfo{year}{1982}).

\bibitem[{\citenamefont{{Kiljunen} et~al.}(2000)\citenamefont{{Kiljunen},
  {Eloranta}, {Kunttu}, {Khriachtchev}, {Pettersson}, and
  {R{\"a}s{\"a}nen}}}]{2000JChPh.112.7475K}
\bibinfo{author}{\bibfnamefont{T.}~\bibnamefont{{Kiljunen}}},
  \bibinfo{author}{\bibfnamefont{J.}~\bibnamefont{{Eloranta}}},
  \bibinfo{author}{\bibfnamefont{H.}~\bibnamefont{{Kunttu}}},
  \bibinfo{author}{\bibfnamefont{L.}~\bibnamefont{{Khriachtchev}}},
  \bibinfo{author}{\bibfnamefont{M.}~\bibnamefont{{Pettersson}}},
  \bibnamefont{and}
  \bibinfo{author}{\bibfnamefont{M.}~\bibnamefont{{R{\"a}s{\"a}nen}}},
  \bibinfo{journal}{\jcp} \textbf{\bibinfo{volume}{112}}, \bibinfo{pages}{7475}
  (\bibinfo{year}{2000}).

\bibitem[{\citenamefont{Peyerimhoff and Buenker}(1982)}]{ChemPhys.72.111}
\bibinfo{author}{\bibfnamefont{S.~D.} \bibnamefont{Peyerimhoff}}
  \bibnamefont{and} \bibinfo{author}{\bibfnamefont{R.~J.}
  \bibnamefont{Buenker}}, \bibinfo{journal}{Chem. Phys.}
  \textbf{\bibinfo{volume}{72}}, \bibinfo{pages}{111} (\bibinfo{year}{1982}).

\bibitem[{\citenamefont{Morrison}(1982)}]{PhysRevA.25.1445}
\bibinfo{author}{\bibfnamefont{M.~A.} \bibnamefont{Morrison}},
  \bibinfo{journal}{Phys. Rev. A} \textbf{\bibinfo{volume}{25}},
  \bibinfo{pages}{1445} (\bibinfo{year}{1982}).

\bibitem[{\citenamefont{{Morgan}}(1998)}]{1998PhRvL..80.1873M}
\bibinfo{author}{\bibfnamefont{L.~A.} \bibnamefont{{Morgan}}},
  \bibinfo{journal}{Phys. Rev. Lett.} \textbf{\bibinfo{volume}{80}},
  \bibinfo{pages}{1873} (\bibinfo{year}{1998}).

\end{thebibliography}

\clearpage


%



\clearpage

\clearpage

\clearpage

\clearpage

\clearpage

\clearpage

%


\begin{table}%
\caption{\label{tab1}
List of target states included in the present R-matrix calculations.
}
\begin{ruledtabular}
\begin{tabular}{cc}
O$_2$  & ${X}^3\Sigma^{-}_{g}$,${a}^1\Delta_{g}$,
${b}^1\Sigma^{+}_{g}$,${c}^1\Sigma^{-}_{u}$,${A'}^3\Delta_{u}$,
${A}^3\Sigma^{+}_{u}$,${B}^3\Sigma^{-}_{u}$,${1}^1\Delta_{u}$,
${f'}^1\Sigma^{+}_{u}$, 
${1}^1\Pi_{g}$,${1}^3\Pi_{g}$,${1}^1\Pi_{u}$, ${1}^3\Pi_{u}$     \\
B$_2$  & ${X}^3\Sigma^{-}_{g}$, ${a}^5\Sigma^{-}_{u}$, ${b}^1\Delta_{g}$, 
${A}^3\Pi_{u}$, ${c}^1\Sigma^{+}_{g}$, ${1}^3\Delta_{u}$,
${1}^1\Pi_{u}$, ${1}^3\Sigma^{+}_{u}$, ${1}^3\Sigma^{-}_{u}$, 
${1}^3\Pi_{g}$, ${2}^1\Sigma^{+}_{g}$, ${1}^1\Sigma^{-}_{u}$, 
${2}^3\Pi_{g}$, ${1}^1\Pi_{g}$   \\ 
S$_2$  &  ${X}^3\Sigma^{-}_{g}$, ${a}^1\Delta_{g}$, ${b}^1\Sigma^{+}_{g}$, 
${c}^1\Sigma^{-}_{u}$, ${A'}^3\Delta_{u}$, ${A}^3\Sigma^{+}_{u}$,
${B'}^3\Pi_{g}$, ${B}^3\Sigma^{-}_{u}$, ${1}^1\Pi_{g}$, 
${1}^1\Delta_{u}$, ${B''}^3\Pi_{u}$, ${1}^1\Sigma^{+}_{u}$, 
${1}^1\Pi_{u}$   \\
Si$_2$  &  ${X}^3\Sigma^{-}_{g}$, ${1}^3\Pi_{u}$, ${1}^1\Delta_{g}$, 
${1}^1\Sigma^{+}_{g}$, ${1}^1\Pi_{u}$, ${2}^1\Sigma^{+}_{g}$, 
${1}^5\Pi_{g}$, ${1}^3\Pi_{g}$, ${1}^3\Sigma^{+}_{u}$, 
${1}^1\Sigma^{-}_{u}$, ${2}^3\Pi_{g}$, ${1}^3\Delta_{u}$, 
${2}^3\Sigma^{+}_{u}$, ${1}^1\Pi_{g}$, ${1}^3\Phi_{g}$  \\ 
\end{tabular}
\end{ruledtabular}
\end{table}

\begin{table}%
\caption{\label{tab2}
Division of the orbital set in each symmetry for e-B$_2$ case. 
}
\begin{ruledtabular}
\begin{tabular}{lrrrrrrrr}
Symmetry  & $A_g$ & $B_{2u}$ & $B_{3u}$ & $B_{1g}$ &  $B_{1u}$ &
   $B_{3g}$ &  $B_{2g}$ &   $A_u$    \\
\hline
Valence   & 1-3$a_g$ & 1$b_{2u}$ &  1$b_{3u}$ &  &  1-3$b_{1u}$ &
  1$b_{3g}$  &  1$b_{2g}$ &   \\ 
Extra virtual   & 4$a_g$ & 2$b_{2u}$ &  2$b_{3u}$ & 1$b_{1g}$ & 
4$b_{1u}$ &  2$b_{3g}$ & 2$b_{2g}$  & 1$a_u$  \\
Continuum & 5-43$a_g$ & 3-23$b_{2u}$ &  3-23$b_{3u}$ & 
2-18$b_{1g}$  &  5-25$b_{1u}$ &  3-20$b_{3g}$ &  
3-20$b_{2g}$ & 2-7$a_u$    \\
\end{tabular}
\end{ruledtabular}
\end{table}

\clearpage

\begin{table}%
\caption{\label{tab3}
 The vertical excitation energies of the first 5 excited states 
 for B$_2$ molecule, with the previous full configuration interaction (FCI)
 results of Hald et al.\cite{2001JChPh.115..671H}. The unit of energy is eV.
}
\begin{ruledtabular}
\begin{tabular}{ccc}
State & This work & FCI \\
\hline
${X}^3 \Sigma^{+}_{-}$  &  0.00  & 0.00  \\
${a}^5 \Sigma_{u}^{-}$  &  0.06  & 0.26  \\
${b}^1 \Delta_{g}$      &  0.71  & 0.63  \\
${A}^3 \Pi_{u}$         &  0.91  & 0.69  \\
${c}^1 \Sigma_{g}^{+}$  &  0.98  & 0.98  \\
${1}^3 \Delta_{u}$      &  1.75  & 1.66  \\
\end{tabular}
\end{ruledtabular}
\end{table}

\begin{table}%
\caption{\label{tab4}
 The vertical excitation energies of the first 5 excited states 
 for S$_2$ molecule, with the previous MRD CI results of Hess et al.\cite{ChemPhys.71.79} 
 and MRCI results of Kiljunen et al.\cite{2000JChPh.112.7475K}. 
 The unit of energy is eV.
}
\begin{ruledtabular}
\begin{tabular}{cccc}
State & This work & Previous MRD CI & Previous MRCI \\
\hline
${X}^3 \Sigma^{+}_{-}$  &  0.00  & 0.00 & 0.00  \\
${a}^1 \Delta_{g}$      &  0.60  & 0.68 & 0.55  \\
${b}^1 \Sigma_{g}^{+}$  &  0.92  & 1.04 & 0.99  \\
${c}^1 \Sigma_{u}^{-}$  &  2.77  &      & 2.45  \\
${A'}^3 \Delta_{u}$     &  2.93  &      & 2.59  \\
${A}^3 \Sigma_{u}^{+}$  &  3.03  &      & 2.58  \\
\end{tabular}
\end{ruledtabular}
\end{table}

\begin{table}%
\caption{\label{tab5}
 The vertical excitation energies of the first 5 excited states 
 for Si$_2$ molecule, with the previous MRD CI results by 
 Peyerimhoff and Buenker \cite{ChemPhys.72.111}. 
 The unit of energy is eV.
}
\begin{ruledtabular}
\begin{tabular}{ccc}
State & This work & Previous calculation \\
\hline
${X}^3 \Sigma^{+}_{-}$  &  0.00  &  0.00 \\
${1}^3 \Pi_{u}$         &  0.22  &  0.07 \\
${1}^1 \Delta_{g}$      &  0.62  &  0.53 \\
${1}^1 \Sigma_{g}^{+}$  &  0.87  &  0.71 \\
${1}^1 \Pi_{u}$         &  0.90  &  0.63 \\
${2}^1 \Sigma_{g}^{+}$  &  1.44  &  1.14 \\
\end{tabular}
\end{ruledtabular}
\end{table}


\end{document}